# Logarithmic Correlators in Non-relativistic Conformal Field Theory

Ali Hosseiny and Shahin Rouhani

Physics Department,

Sharif University of Technology,

Tehran PO Box 11165-9161, Iran.

Email: alihosseiny@physics.sharif.edu , srouhani@sharif.ir

## Abstract

We show how logarithmic terms may arise in the correlators of fields which belong to the representation of the Schrödinger-Virasoro algebra (SV) or the affine Galilean Conformal Algebra (GCA). We show that in GCA, only scaling operator can have a Jordan form and rapidity cannot. We observe that in both algebras logarithmic dependence appears along the time direction alone.

## 1-Introduction

Non-relativistic conformal invariance has recently received a lot of attention in the context of AdS/CFT correspondence [1-6]. Historically though there were other reasons; on the one hand investigations of the symmetries of Schrödinger equation [7-10] on the other hand, the application to hydrodynamics [as an example see 11]. The recent activity in AdS/CFT has pointed towards the application of this correspondence to condensed matter setting [4,6], in particular it has been observed that asymptotic symmetry algebra of geometries with Schrodinger isometry in any dimension is an infinite dimensional algebra containing one copy of Virasoro [12,13].

Though being nonrelativistic, Schrodinger equation in addition to Galilean symmetry carries two more symmetries. One is inversion:

$$K = -(tx_i\partial_i + t^2\partial_t) \ , \tag{1.1}$$

which corresponds to the special conformal transformation (SCT). The other symmetry of Schrodinger equation is scaling:

$$D = -(2t\partial_t + x_i\partial_i) \ . \tag{1.2}$$

Having scale invariance in Schrodinger equation can signal that there might exist some sort of Nonrelativistic conformal invariance under additional conditions. Historically Conformal Field Theory (CFT) was developed to address the quantum mechanics of string theory, thus it was developed in a relativistic setting. Later it was realised that CFT may be used to explain physical systems at their critical points [14,15]. Such systems are often at low temperatures and thus nonrelativistic, but since they are studied at thermal equilibrium time does not play a role and a relativistic symmetry finds application. However, when we look at the dynamics of systems close to their critical point, non relativistic CFT (NRCFT) becomes relevant [see for example 16].

Beside Schrodinger algebra, there is another Galilean algebra investigated by physicists; the Galilean Conformal Algebra (GCA) [17]. This algebra is obtained via a direct contraction of relativistic conformal algebra. In this contraction one investigates the behaviour of CFT for low velocities or:

$$x \to x/c, \qquad t \to t; \qquad c \to \infty \ . \tag{1.3}$$

As well as the affine Schrödinger symmetry (SV) one may have the affine Galilean Conformal Algebra (GCA) [18,19,20] which can be obtained via a direct contraction of relativistic conformal symmetry [21]. GCA has been proposed as a symmetry of nonrelativistic incompressible fluid dynamics [22], but this possibility has been challenged [23,24] . For a study of differential equations which admit this symmetry see [24]. Mass central charge is the most common charge and it results Bargmann superselection rules for Schrodinger correlation functions. However in 2+1 another central charge, named "Exotic" is of interest [25-28].

Looking for other non relativistic conformal symmetries, it was observed that the class of nonrelativistic conformal algebras are quite restricted [19,29]. This class scales space and time anisotropically:

$$t \to \lambda t, \qquad \vec{x} \to \lambda^l \vec{x} . \tag{1.4}$$

Each algebra of this class is characterized by a fraction $l$ which is one over the dynamical exponent. Closure of the algebra forces $l$ to be half integer. However in the special case of $d=2+1$ another class of Galilei symmetries becomes relevant for nonrelativistic systems without any restrictions on the dynamical index [21]. This is directly related to the fact that in $d=2+1$ we have a Virasoro symmetry along the spatial dimensions. Therefore the requirement of closure may be relaxed. Clearly a wealth of non relativistic conformal symmetries in $d+1$ dimension has been investigated, but a systematic classification has yet to come.

It is natural to wonder about the quantum realisations of these symmetries, their representations and correlation functions. Some attempts have been made at constructing NRCFT's [4, 16, 30-38]. Many interesting features have been discovered and many questions remain. In this paper we look at the question of whether logarithmic correlators may appear in NRCFT's analogous to their relativistic counter parts [39]. Logarithmic Conformal Field Theories (LCFT's) arise when the action of $L_0$ on primary fields is not diagonal. This may happen in some ghost theories such as the $c=-2$ theory [40, 41]. Generically LCFT's are non unitary theories; however applications of LCFT's to some statistical models have been suggested [42-47]. For reviews of LCFT's see [48,49]. More recently LCFT's have become relevant in the context of Topologically Massive Gravity [50-52].

This paper is organised as follows. In section 2 we review logarithmic CFT. In section 3 we review Schrodinger algebra, its representations, correlation functions and null vectors [53].We then obtain its correlation functions for a Jordan action of the scaling operator which would be logarithmic. In section 4 we review GCA, its representations and correlation functions in 1+1 dimensions and investigate its logarithmic correlation functions. In section 5 we close by some concluding remarks.

## 2. Logarithmic Conformal Field Theories (LCFT's)

Logarithmic conformal field theories are based on indecomposable but reducible representations of the Virasoro algebra. Let us consider the case where the generators of $L^0$ can be transformed into Jordan normal form. A rank $r$ Jordan cell is spanned by $r$ states

$$|h,k\rangle, \quad k=0..r-1 ,$$

$$L^0|h,k\rangle = h|h,k\rangle + (1-\delta_{k,0})|h,k-1\rangle , \tag{2.1}$$

where $h$ is the conformal weight. The parameter k grades the fields within the Jordan cell and will be referred to as Jordan level. In this section we shall concentrate on level 2, which is indeed the only level with a particular field theoretic example of *c=-2* ghost theory. The equation (2.1) may be rewritten using a nilpotent variable [54] in a much more convenient way (*r=2*):

$$\theta^2 = 0 \,,$$

$$L^0|h+\theta\rangle = (h+\theta)|h+\theta\rangle \,, \quad (2.2)$$

attaching the interpretation that the highest weight has a graded eigenvalue. To connect up with (2.1) let us note that a Taylor expansion of the state $|h+\theta\rangle$ has only two terms:

$$|h+\theta\rangle = |h,0\rangle + \theta|h,1\rangle \,. \quad (2.3)$$

The above expression for a Jordan cell of rank 2 immediately results. This extension immediately results in logarithmic terms appearing in the correlators. The state operator correspondence requires two fields to exist with action of $L^0$ defined from (2.1):

$$L^0\phi_h(z)|0\rangle = h\phi_h(z)|0\rangle \quad ,$$

$$L^0\psi_h(z)|0\rangle = h\psi_h(z)|0\rangle + \phi_h(z)|0\rangle \,, \quad (2.4)$$

where $|0\rangle$ is the usual invariant vacuum. The pair of fields $\psi, \phi$ are logarithmic partners, with the same conformal weight. In the minimal series each highest weight uniquely determines a primary field but here clearly we have degeneracy. Following the notion of nilpotent variables we can also define a field with nilpotent variables:

$$\mathbf{\Phi}(z,\theta) = \phi(z) + \theta\psi(z) \,, \quad (2.5)$$

which is a field, with a conformal weight with a nilpotent component $h+\theta$. This implies that under a mapping of the complex plane $z \to \omega(z)$ we have:

$$\mathbf{\Phi}(z,\theta) \to (\frac{\partial\omega}{\partial z})^{\boldsymbol{h+\theta}} \, \mathbf{\Phi}(\omega,\theta) \,, \quad (2.6)$$

resulting in the usual transformation laws for the primary field $\phi(z)$, but logarithmic transformation law for the partner field is:

$$\psi_h(z) \to (\frac{\partial\omega}{\partial z})^{\boldsymbol{h}}(\psi_h(z) + \log\,(\frac{\partial w}{\partial z})\phi(z)) \,. \quad (2.7)$$

We can now ask what the correlators will have to be if they are to respect the transformation laws of a logarithmic pair. As before, we can proceed by requiring invariance under the corresponding generators of the Mobius group, leading to the correlators

$$\langle\phi|\phi\rangle = 0 \,,$$

$$\langle\phi|\psi\rangle = az^{-2h} \,,$$

$$\langle\psi|\psi\rangle = z^{-2h}(b - 2a\,log(z)) \,. \quad (2.8)$$

We observe the unusual property that the correlator $\langle\phi|\phi\rangle$ vanishes. This has the following interpretation. Since OPE requires the identity to appear in the rhs of the product of a field with itself, the only possibility is that the identity has vanishing expectation value. The resolution of this apparently conflicting result comes by noting that an LCFT may have a distinguished operator pair; the identity and its logarithmic partner:

$$I(\theta) = I + \theta\Omega \ , \tag{2.9}$$

with correlators:

$$\langle I\rangle = 0 \ ,$$
$$\langle\Omega\rangle = 1 \ ,$$
$$\langle\Omega(0)\Omega(z)\rangle = \log{(z)} \ , \tag{2.10}$$

which is consistent with zero conformal weight but unusual in giving zero expectation for the identity and the existence of a partner for the identity which has some kind of z dependence. Clearly the state-operator correspondence implies that there should exist two candidates for the vacuum, one for each identity. This also means that the other distinguished operator within the conformal algebra i.e. the energy momentum tensor must also have a logarithmic partner:

$$\begin{aligned} T(z,\theta) &= L^{-2}I(\theta) \\ &= T(z) + \theta\ t(z) \ , \end{aligned} \tag{2.11}$$

the second entity i.e. *t(z)* is observed within the $AdS_3/LCFT_2$ correspondence [50-52]. The requirements of invariance under conformal Ward identity produces logarithmic terms. However let us note that, not all primary fields may have Jordan cell structure. It is possible that the identity does not have a Jordan cell but other primaries have, in such a case the compatibility of the vanishing n-point functions and OPE needs an explanation.

Now we consider the appearance of these Jordan cells in the non relativistic cases i.e. the Schrodinger algebra and the GCA.

## 3. Schrodinger Algebra

Schrodinger algebra is the symmetry of Schrodinger equation [7,8]:

$$P_i = \partial_i, \qquad H = -\partial_t, \qquad B_i = t\partial_i,$$
$$J_{ij} = -(x_i\partial_j - x_j\partial_i) \ , \qquad D = -(2t\partial_t + x_i\partial_i),$$
$$K = -(tx_i\partial_i + t^2\partial_t) \ . \tag{3.1}$$

These generators produce Schrodinger group which is the group of the following transformations:

$$\vec{r} \to \overrightarrow{r'} = \frac{\mathcal{R}\vec{r}+\vec{V}t+\vec{a}}{\gamma t+\delta}, \quad t \to t' = \frac{\alpha t+\beta}{rt+\delta}, \qquad \alpha\delta - \beta\gamma = 1 . \tag{3.2}$$

This algebra admits mass central charge

$$[B_i, P_j] = m\delta_{ij} . \tag{3.3}$$

Non-vanishing commutators are:

$$[H, B_i] = -P_i, [K, P_i] = B_i, \quad [H, K] = D, \quad [D, B_i] = -B_i,$$

$$[D, P_i] = P_i, \quad [D, K] = -2K, \qquad [D, H] = H$$

$$[J_{ij}, J_{kl}] = SO(d), \qquad [J_{ij}, B_k] = -(B_i\delta_{jk} - B_j\delta_{ik})$$

$$[J_{ij}, P_k] = -(P_i\delta_{jk} - P_j\delta_{ik}) . \tag{3.4}$$

*H,D* and *K* form an *SL(2,R)*. It suggests that this algebra may have a Virasoro-like affine extension. It does [16] and is called Schrodinger Virasoro algebra (SV):

$$T^n = -t^{n+1}\partial_t - \frac{1}{2}(n+1)t^n x_i\partial_i - \frac{1}{4}n(n+1)\mathcal{M}t^{n-1}x^2,$$

$$P^m{}_i = -t^{m+\frac{1}{2}}\partial_i - (m+\frac{1}{2})t^{m-1/2}x_i\mathcal{M},$$

$$M^n = -\mathcal{M}t^n \ , \tag{3.5}$$

where $n \in \mathbb{Z}$ and $m \in \mathbb{Z} + \frac{1}{2}$. In this format Schrodinger algebra would be:

$$P^{-\frac{1}{2}}{}_i = -\partial_i = -P_i, \quad P^{\frac{1}{2}}{}_i = B_i,$$

$$T^{-1} = H, \qquad T^0 = D/2, \qquad T^1 = K, \tag{3.6}$$

and the commutators in 1+1 are

$$[T^n, T^m] = (n-m)T^{n+m}, \qquad [T^n, P^m] = (\frac{n}{2} - m)P^{n+m} ,$$

$$[T^n, M^m] = -mM^{n+m} , \qquad [P^n, P^m] = (n-m)M^{n+m},$$

$$[P^n, M^m] = [M^n, M^m] = 0 . \tag{3.7}$$

Before going on to more details we would like to recall that to have a proper form for invariance under this algebra we need to define another parameter ξ and wave functions $\psi(\xi, x, t)$ as [38]:

$$\phi(x,t) = \frac{1}{\sqrt{2\pi}}\int_{\mathbb{R}} d\xi \exp(-im\xi)\, \psi(\xi, x, t) . \tag{3.8}$$

Now, Schrodinger equation turns to a Klein-Gordon form:

$$(-2\partial_\zeta\partial_t + \partial_i\partial_i)\psi = 0 , \tag{3.9}$$

and the metric in this space is:

$$ds^2 = 2d\zeta dt + dx.dx \; . \tag{3.10}$$

The algebraic elements can be rewritten and dealt with in this new coordinates. However, we expect the final answer not to change. The only correction for two-point functions would be a Heaviside function of time. To see this effect and some other results see [31].

To review representations and correlation functions of Schrodinger algebra we follow [16,30,31]. Representations are in 1+1. However, two-point functions work for any dimensions. In Schrodinger symmetry scaling fields are recognized by their scaling weight *h* and mass $\mathcal{M}$:

$$[T^0, \phi] = h\phi,$$

$$[M^0, \phi] = \mathcal{M}\phi \; . \tag{3.11}$$

From equation (3.7) it can be easily seen that $T^n, M^n, P^m$ lower *h* for *n, m>0* and raise it for *n,m<0*. In other words

$$[T^0, [T^n, \phi]] = (h - n)[T^n, \phi]$$

$$[T^0, [P^m, \phi]] = (h - m)[P^m, \phi]$$

$$[T^0, [M^n, \phi]] = (h - n)[M^n, \phi] \; . \tag{3.12}$$

Since $M^0$ as a central charge commutes with all operators, neither of them can lower nor raise the mass $\mathcal{M}$. Now, considering a lowest primary operator we can build the descendant operators. We consider a primary operator as the lowest bound which is annihilated by all lowering operators or:

$$[T^n, \phi] = 0 \; , \qquad [P^m, \phi] = 0 \; , \qquad [M^n, \phi] = 0 \tag{3.13}$$

for any *n ,m >0*. To go further, borrowing state-operator correspondence we represent operators by their corresponding states $|h\rangle$ . So, for a state with scaling weight of *h* and mass $\mathcal{M}$ we have

$$T^o|h\rangle_{\mathcal{M}} = h|h\rangle_{\mathcal{M}} \; ,$$

$$M^0|h\rangle_{\mathcal{M}} = \mathcal{M}|h\rangle_{\mathcal{M}} \; . \tag{3.14}$$

However, since $\mathcal{M}$ is central charge and won't be changed by operators we represent states only by *h*. As mentioned, a tower of representations is built by acting raising operators on the lowest state:

$$T^{-n}|h\rangle \rightarrow \; |h + n\rangle \; ,$$

$$P^{-m}|h\rangle \; \rightarrow |h + m\rangle,$$

$$M^{-n}|h\rangle \to |h+n\rangle\,, \qquad\qquad n,\, m < 0\,. \tag{3.15}$$

The first level is obtained by acting $P^{-1/2}$ on lowest state or $P^{-1/2}|h\rangle$. The second level can be obtained by acting either $P^{-1/2} * P^{-1/2}$ or $T^{-1}$ on $|h\rangle$. However these two states will be dependent for $h = 1/4$ and the first null vector shows up:

$$|\chi\rangle = ((P^{-\frac{1}{2}})^2 - 2\mathcal{M}T^{-1})|h\rangle\,. \tag{3.16}$$

But this is exactly the Schrodinger equation. The next null vector is obtained in the third level for $h = 11/12$ namely

$$|\chi\rangle = (3T^{-1}P^{-\frac{1}{2}} - 2P^{-\frac{3}{2}} + \frac{3}{2\mathcal{M}}(P^{-\frac{1}{2}})^3)|h\rangle\,. \tag{3.17}$$

It leads to another differential equation namely:

$$\left(3\partial_t\partial_x - 2\left(\frac{\partial_x}{t} - \frac{\mathcal{M}}{t^2}x\right) - \frac{3}{2\mathcal{M}}\partial_x^3\right)\psi = 0 \quad , \tag{3.18}$$

which is invariant under scaling [53].

For obtaining two-point functions of primary fields we look at the infinitesimal action of Schrodinger Virasoro elements on them:

$$[T^n, \phi(x,t)] = (t^{n+1}\partial_t + \frac{n+1}{2}t^n x.\partial_x + \frac{n(n+1)}{4}\mathcal{M}t^{n-1}x^2 + (n+1)ht^n)\phi(x,t)\,,$$

$$[P_i^m, \phi(x,t)] = (t^{m+\frac{1}{2}}\partial_i + \left(m+\frac{1}{2}\right)\mathcal{M}t^{m-\frac{1}{2}}x_i)\phi(x,t),$$

$$[M^n, \phi(x,t)] = (\mathcal{M}t^n\phi(x,t))\,. \tag{3.19}$$

The quasi-primary fields transfer as above as well under Schrodinger algebra ;$\{T^{-1}, T^0, T^1, P^{-\frac{1}{2}}, P^{\frac{1}{2}}, M^0\}$.

Before going to more details we need to note that Schrodinger action is:

$$S = \int dtdx[\mathcal{M}(\phi^*\partial_t\phi - \phi\partial_t\phi^*) + \partial_x\phi\partial_x\phi^*]\,, \tag{3.20}$$

and the equations of motion are:

$$2\mathcal{M}\partial_t\phi - \partial_x\partial_x\phi = 0\,,$$

$$2\mathcal{M}\partial_t\phi^* + \partial_x\partial_x\phi^* = 0\,. \tag{3.21}$$

So, for any field $\phi$ with mass $\mathcal{M}$ there is a conjugate field $\phi^*$, with mass -$\mathcal{M}$ and in equations (3.19) $\mathcal{M}$ should appear with opposite sign for conjugate fields. Now we follow up to find two-point functions of these fields. For two-point functions of quasi-primary fields we suppose:

$$F = F\big(x_1, x_{2;}t_1, t_2\big) = \langle \phi_1(x_1, t_1)\phi_2^*(x_2, t_2)\rangle \, . \qquad (3.22)$$

Invariance under translation $P^{-1/2}$and energy conservation $T^{-1}$result in:

$$F = F(x,t) \, , \qquad \text{where} \qquad x = x_1 - x_2 \, , \qquad t = t_1 - t_2 \, . \qquad (3.23)$$

Invariance under $P^{1/2}$ appears as:

$$\big(t_1\partial_{x_1} + \mathcal{M}_1 x_1 + t_2\partial_{x_2} - \mathcal{M}_2 x_2\big)F(x,t) = 0 \, . \qquad (3.24)$$

It would be possible if and only if we have:

$$\mathcal{M}_1 = \mathcal{M}_2 \, ,$$

$$F(x,t) = u(t)\exp\big(-\tfrac{\mathcal{M}_1 x^2}{2t}\big) \, . \qquad (3.25)$$

Now, we desire scaling behaviour under $T^0$ transformation:

$$\left(t\partial_t + \tfrac{1}{2}x\partial_x + h\right)F(x,t) = 0 \, , \qquad (3.26)$$

where $h = h_1 + h_2$. Putting *F* from (3.25) we get an equation for $u(t)$

$$t\dot{u}(t) + hu(t) = 0 \, , \qquad (3.27)$$

or

$$F(x,t) = at^{-h}\delta_{\mathcal{M}_1,\mathcal{M}_2}\exp\big(-\tfrac{\mathcal{M}_1 x^2}{2t}\big) \, . \qquad (3.28)$$

Desiring invariance under $T^1$ results in having $h_1 = h_2$. So, the final answer is:

$$F(x,t) = at^{-2h_1}\delta_{h_1,h_2}\delta_{\mathcal{M}_1,\mathcal{M}_2}\exp\big(-\tfrac{\mathcal{M}_1 x^2}{2t}\big) \, . \qquad (3.29)$$

As mentioned before, following the covariant form of Schrodinger equation results in a correction for this answer [31]:

$$F(x,t) = at^{-2h_1}\Theta(t_1 - t_2)\delta_{h_1,h_2}\delta_{\mathcal{M}_1,\mathcal{M}_2}\exp\big(-\tfrac{\mathcal{M}_1 x^2}{2t}\big) \qquad , \qquad (3.30)$$

where Θ is the Heaviside function.

To investigate the logarithmic structure in the representations of this algebra we start by a Jordan form for our scaling operator:

$$T^0\phi_h(z)|0\rangle = h\phi_h(z)|0\rangle \, ,$$

$$T^0\psi_h(z)|0\rangle = h\psi_h(z)|0\rangle + \phi_h(z)|0\rangle \, . \qquad (3.31)$$

These equations may be summed up in the nilpotent formalism [54] as:

$$\mathbf{\Phi}(z,\theta) = \phi(z) + \theta\psi(z) \, ,$$

$$\mathbf{\Phi}(z,\theta)|0\rangle = |h+\theta\rangle \,,$$

$$T^0|h+\theta\rangle = (h+\theta)|h+\theta\rangle \,. \qquad (3.32)$$

Now we define two-point correlation function of these fields as:

$$F = F(x_1, x_2; t_1, t_2; \theta_1, \bar{\theta}_2) = \langle \mathbf{\Phi}_1(x_1, t_1, \theta_1)\mathbf{\Phi}_2^*(x_2, t_2, \bar{\theta}_2)\rangle \,. \qquad (3.33)$$

From translation symmetry in space and time we have $F$ as a function of $x$, $t$ .Invariance under $P^{1/2}$ results in having the form:

$$F(x_1, x_2; t_1, t_2; \theta_1, \bar{\theta}_2) = u(t, \theta_1, \bar{\theta}_2)\delta_{\mathcal{M}_1\mathcal{M}_2}\exp\left(-\frac{\mathcal{M}_1 x^2}{2t}\right). \qquad (3.34)$$

Now, we desire invariance under $T^0$ which appears as

$$\left(t\partial_t + \frac{1}{2}x\partial_x + h_1 + h_2 + \theta_1 + \bar{\theta}_2\right)F(x, t, \theta_1, \bar{\theta}_2) = 0 \,, \qquad (3.35)$$

Note that here a conjugate nilpotent variable $\bar{\theta}_2$ has appeared due to presence of the conjugate field $\mathbf{\Phi}_2^*$. Equation 3.35 restricts $F$ to be:

$$F = t^{-(h_1+h_2+\theta_1+\bar{\theta}_2)}(a + b(\theta_1 + \bar{\theta}_2) + c\theta_1\bar{\theta}_2)\,\delta_{\mathcal{M}_1,\mathcal{M}_2}\exp\left(-\frac{\mathcal{M}_1 x^2}{2t}\right). \qquad (3.36)$$

The final format is obtained by imposing invariance under $T^1$ :

$$\left(t_1^2\partial_{t_1} + t_1x_1\partial_{x_1} + \frac{\mathcal{M}_1}{2}x_1^2 + 2(h_1+\theta_1)t_1 + t_2^2\partial_{t_2} + t_2x_2\partial_{x_2} - \frac{\mathcal{M}_2}{2}x_2^2 + 2(h_2 + \bar{\theta}_2)t_2\right)F(x, t, \theta_1, \bar{\theta}_2) = 0 \,, \qquad (3.37)$$

and would be:

$$F = t^{-2h_1-\theta_1-\bar{\theta}_2}\delta_{h_1,h_2}\delta_{\mathcal{M}_1,\mathcal{M}_2}\exp\left(-\frac{\mathcal{M}_1 x^2}{2t}\right)(b(\theta_1+\bar{\theta}_2) + c\theta_1\bar{\theta}_2)$$

$$= t^{-2h_1}\delta_{h_1,h_2}\delta_{\mathcal{M}_1,\mathcal{M}_2}\exp\left(-\frac{\mathcal{M}_1 x^2}{2t}\right)(b(\theta_1+\bar{\theta}_2) + \theta_1\bar{\theta}_2(c - 2b\,logt)) \,. \qquad (3.38)$$

If we compare this expansion by an expansion of correlation function in (3.33) we obtain two-pint correlations of all fields:

$$\langle\phi_1(x_1, t_1)\phi_2^*(x_2, t_2)\rangle = 0 \,,$$

$$\langle\phi_1(x_1, t_1)\psi_2^*(x_2, t_2)\rangle = bt^{-2h_1}\delta_{\mathcal{M}_1,\mathcal{M}_2}\exp\left(-\frac{\mathcal{M}_1 x^2}{2t}\right),$$

$$\langle\psi_1(x_1, t_1)\psi_2^*(x_2, t_2)\rangle = t^{-2h_1}\delta_{h_1,h_2}\delta_{\mathcal{M}_1,\mathcal{M}_2}\exp\left(-\frac{\mathcal{M}_1 x^2}{2t}\right)(c - 2b\,log(t)) \,. \qquad (3.39)$$

## 4. Galilean Conformal Algebra (GCA)

Galilean conformal algebra [17] can be obtained directly from contraction. However as it was observed in [19,29], it corresponds to $l=1$ of $l$-Galilei[1] algebra. In d+1, in addition to Galilean algebra,$(J_{i,j}, H, P_i, B_i)$, it consist of *d+2* more elements:

$$D = -(t\partial_t + x_i\partial_i)\,, \qquad K = K_0 = -(2tx_i\partial_i + t^2\partial_t)\,,$$

$$K_i = t^2\partial_i\,. \tag{4.1}$$

Similar to Schrodinger algebra, this algebra admits an affine extension as well:

$$T^n = -(n+1)t^n x_i\partial_i - t^{n+1}\partial_t,$$

$$M_i^n = t^{n+1}\partial_i,$$

$$J_{ij}^n = -t^n(x_i\partial_j - x_j\partial_i)\,, \tag{4.2}$$

in which the central algebra is

$$T^{-1} = H, \qquad T^0 = D, \qquad T^1 = K\,,$$

$$M_i^{-1} = P_i, \qquad M_i^0 = B_i, \qquad M_i^1 = K_i\,, \tag{4.3}$$

and commutation relations are:

$$[T^m, T^n] = (m-n)T^{m+n}, \qquad [T^m, J_a^n] = -nJ_a^{m+n},$$

$$[J_a^m, J_b^n] = f_{abc}J_c^{m+n}\,, \qquad [T^m, M_i^n] = (m-n)M_i^{m+n},$$

$$\left[M_i^m, M_j^n\right] = 0, \qquad \left[M_i^m, J_{jk}^n\right] = (M_j^{m+n}\delta_{ik} - M_k^{m+n}\delta_{ij})\,. \tag{4.4}$$

$f_{abc}$ are SO(d) structure constants.

Correlation functions of this algebra were first given in [32] and then rederived in [35,33]. To obtain representations in 1+1 we first observe that full GCA in this dimension can be obtained directly from a contraction of $CFT_2$. To observe this contraction we go to complex coordinates:

$$z = t + i\,x\,, \qquad \bar{z} = t - i\,x\,. \tag{4.5}$$

Relativistic conformal symmetry of *d=2* contains two Virasoro

$$L^n = -z^{n+1}\partial_z\,, \qquad \bar{L}^n = -\bar{z}^{n+1}\partial_{\bar{z}}\,, \tag{4.6}$$

which are the generators of holomorphic and antiholomorphic transformations. Now, we try to impose contraction

$$x \to \frac{x}{c}\,, \qquad t \to t\,, \qquad c \to \infty\,, \tag{4.7}$$

[1] These algebras sometimes have been referred to as spin-*l* Galilei algebras in the literature but we find this name confusing since *l* has noting to do with spin.

and obtain full GCA form CFT elements [21]:

$$L^n = -\frac{1}{2}(t + i\frac{x}{c})^{n+1}(\partial_t - ic\partial_x)$$

$$= -t^{n+1}(-ic\partial_x + \partial_t + (n+1)\frac{x}{t}\partial_x + O(\frac{1}{c})$$

$$\bar{L}^n = -t^{n+1}(ic\partial_x + \partial_t + (n+1)\frac{x}{t}\partial_x + O(\frac{1}{c}) \,. \tag{4.8}$$

As a limit for $c \to \infty$ we have

$$T^n = L^n + \bar{L}^n + O(\frac{1}{c})\,, \qquad M^n = -i\frac{L^n - \bar{L}^n}{c} + O(\frac{1}{c})\,. \tag{4.9}$$

Now that we have full GCA obtained from contraction of CFT elements, we might be able to obtain its representations by contraction as well. Usually when an algebra is contracted, there is no necessity that its representation are obtained from contraction as well. However, as we will see this contraction represents the algebra and correlation functions properly. For $CFT_2$ representations are built on eigenvectors of $\bar{L}^0$ and $L^0$ :

$$L^0|h,\bar{h}\rangle = h|h,\bar{h}\rangle\,, \qquad \bar{L}^0|h,\bar{h}\rangle = \bar{h}|h,\bar{h}\rangle\,. \tag{4.10}$$

Since in GCA $T^0$ and $M^0$ commute, we investigate if we can make representations based on their eigenvectors and we observe:

$$T^0|h,\bar{h}\rangle = (L^0 + \bar{L}^0)|h,\bar{h}\rangle = (h + \bar{h})|h,\bar{h}\rangle\,,$$

$$M^0|h,\bar{h}\rangle = -\frac{i}{c}(L^0 - \bar{L}^0)|h,\bar{h}\rangle = \frac{i}{c}(\bar{h} - h)|h,\bar{h}\rangle\,.$$

So, if we represent our states by $|\Delta,\xi\rangle$, we have:

$$T^0|\Delta,\xi\rangle = \Delta|\Delta,\xi\rangle\,,$$

$$M^0|\Delta,\xi\rangle = \xi|\Delta,\xi\rangle, \tag{4.11}$$

in which

$$h = \frac{\Delta + i\,c\,\xi}{2}$$

$$\bar{h} = \frac{\Delta - i\,c\,\xi}{2} \tag{4.12}$$

So, we observe that in the contraction limit $h$ and $\bar{h}$ get relatively large. To have them real numbers, in our notation we must have rapidity ξ an imaginary number. If we write $CFT_2$ elements with central charge:

$$[L_m, L_n] = (m-n)L_{m+n} + \frac{1}{12}C_{cft}m(m^2-1)\delta_{m+n,0}\,,$$

$$[\bar{L}_m, \bar{L}_n] = (m-n)\bar{L}_{m+n} + \frac{1}{12}\bar{C}_{cft} m(m^2-1)\delta_{m+n,0} \,. \tag{4.13}$$

then we obtain 2 central charges for GCA

$$[T^m, T^n] = (m-n)T^{m+n} + \frac{1}{12}C_{GT} m(m^2-1)\delta_{m+n,0} \,,$$

$$[T^m, M^n] = (m-n)M^{m+n} + \frac{1}{12}C_{GM} m(m^2-1)\delta_{m+n,0} \,, \tag{4.14}$$

In which we have:

$$C_{cft} = \frac{1}{2}(C_{GT} + icC_{GM})$$

$$\bar{C}_{cft} = \frac{1}{2}(C_{GT} - icC_{GM}) \tag{4.15}$$

From GCA algebra (4.4), it can be observed that for GCA states (4.11), $T^m$ and $M^m$ can act as raising and lowering operators of the scaling weight $\Delta$:

$$T^0 T^n|\Delta,\xi\rangle = (\Delta - n)T^n|\Delta,\xi\rangle \,,$$

$$T^0 M^n|\Delta,\xi\rangle = (\Delta - n)M^n|\Delta,\xi\rangle \,. \tag{4.16}$$

However, for rapidity things are different and we have:

$$M^0 T^n|\Delta,\xi\rangle = \xi T^n|\Delta,\xi\rangle - nM^n|\Delta,\xi\rangle \tag{4.17}$$

Now we try to observe how our states represent these equations. In $CFT_2$ the lowest state is a state which is annihilated by all lowering operators. The other states are obtained by acting raising operators:

$$L^{-m}|h,\bar{h}\rangle = \sqrt{2mh + \frac{1}{12}C_{cft} m(m^2-1)}\,|h+m,\bar{h}\rangle =$$

$$a_m|h+m,\bar{h}\rangle, \qquad \text{m>0} \tag{4.18}$$

In contraction language, it means that a lowest bound state is annihilated by all lowering operators or $T^m$ and $M^m$ for *m>0*. The other states are created by acting raising operators:

$$T^{-m}|\Delta,\xi\rangle = (L^{-m} + \bar{L}^{-m})|h,\bar{h}\rangle = a_m|h+m,\bar{h}\rangle + \bar{a}_m|h,\bar{h}+m\rangle$$

$$= a_m\left|\Delta+m, \xi + i\frac{m}{c}\right\rangle + \bar{a}_m\left|\Delta+m, \xi - i\frac{m}{c}\right\rangle \tag{4.19}$$

It should be noticed that though rapidity change is of the order of $O(\frac{1}{c})$, it should be kept for the sake of having some coefficients such as $a_m\bar{a}_m$ to order of *c*. The action of $M^{-m}$on states would be:

$$M^{-m}|\Delta,\xi\rangle = -\frac{i}{c}(L^{-m} - \bar{L}^{-m})|h,\bar{h}\rangle = -\frac{i}{c}a_m|h+m,\bar{h}\rangle + \frac{i}{c}\bar{a}_m|h,\bar{h}+m\rangle$$

$$= -\frac{i}{c} a_m \left|\varDelta + m, \xi + i\frac{m}{c}\right\rangle + \frac{i}{c} \bar{a}_m \left|\varDelta + m, \xi - i\frac{m}{c}\right\rangle \tag{4.20}$$

Now, we check equation (4.17) and observe that correction for ξ comes handy:

$$M^0 T^{-m}|\Delta, \xi\rangle = M^0 a_m \left|\varDelta + m, \xi + i\frac{m}{c}\right\rangle + M^0 \bar{a}_m \left|\varDelta + m, \xi - i\frac{m}{c}\right\rangle$$

$$= \xi \left(a_m \left|\varDelta + m, \xi + i\frac{m}{c}\right\rangle + \bar{a}_m \left|\varDelta + m, \xi - i\frac{m}{c}\right\rangle\right)$$

$$+ i\frac{m}{c}\left(a_m \left|\varDelta + m, \xi + i\frac{m}{c}\right\rangle - \bar{a}_m \left|\varDelta + m, \xi - i\frac{m}{c}\right\rangle\right)$$

$$= \xi T^n|\Delta, \xi\rangle - mM^n|\Delta, \xi\rangle \tag{4.21}$$

Tow-point functions of GCA as well can be obtained via contraction:

$$\langle\phi_1(x_1, t_1)\phi_2(x_2, t_2)\rangle_{GCA} = lim_{c\to\infty}\langle\phi_1(x_1, t_1)\phi_2(x_2, t_2)\rangle_{CFT}$$

$$= lim_{c\to\infty}\, \delta_{h_1,h_2}\delta_{\bar{h}_1,\bar{h}_2} {z_{12}}^{-2h_1} {\bar{z}_{12}}^{-2\bar{h}_1}$$

$$= lim_{c\to\infty}\, A\delta_{h_1,h_2}\delta_{\bar{h}_1,\bar{h}_2}\left(t_{12} + i\,\frac{x_{12}}{c}\right)^{-(\Delta + i\,c\,\xi)}\left(t_{12} - i\frac{x_{12}}{c}\right)^{-(\Delta - i\,c\,\xi)}$$

$$= a\delta_{\Delta_1,\Delta_2}\delta_{\xi_1,\xi_2} {t_{12}}^{-2\Delta_1} exp\left(\frac{2\xi_1 x_{12}}{t_{12}}\right). \tag{4.22}$$

All these results have been independently obtained in [34]. Now we consider logarithmic representations and find its contracted form. In CFT we have:

$$L^0\left|h, 0, \bar{h}, 0\right\rangle = h\left|h, 0, \bar{h}, 0\right\rangle,$$

$$L^0\left|h, 1, \bar{h}, 0\right\rangle = h\left|h, 1, \bar{h}, 0\right\rangle + \left|h, 0, \bar{h}, 0\right\rangle. \tag{4.23}$$

Now, look at equation (4.11) and contraction (4.9). For the contraction limit we define:

$$|\varDelta, \xi, 0\rangle = \left|h, 0, \bar{h}, 0\right\rangle,$$

$$|\varDelta, \xi, 1\rangle = \left|h, 1, \bar{h}, 0\right\rangle, \tag{4.24}$$

Then, for our scaling operator we have:

$$T^0|\varDelta, \xi, 0\rangle = T^0\left|h, 0, \bar{h}, 0\right\rangle = \varDelta|\varDelta, \xi, 0\rangle,$$

$$T^0|\varDelta, \xi, 1\rangle = T^0\left|h, 1, \bar{h}, 0\right\rangle = h\left|h, 1, \bar{h}, 0\right\rangle + \bar{h}\left|h, 1, \bar{h}, 0\right\rangle + \left|h, 0, \bar{h}, 0\right\rangle$$

$$= \varDelta|\varDelta, \xi, 1\rangle + |\varDelta, \xi, 0\rangle. \tag{4.25}$$

And for rapidity operator:

$$M^0|\varDelta, \xi, 0\rangle = M^0\left|h, 0, \bar{h}, 0\right\rangle = -i\frac{h}{c}\left|h, 0, \bar{h}, 0\right\rangle + i\frac{\bar{h}}{c}\left|h, 0, \bar{h}, 0\right\rangle = \xi|\varDelta, \xi, 0\rangle,$$

$$M^0|\Delta,\xi,1\rangle = M^0\big|h,1,\bar{h},0\rangle = -i\frac{h}{c}\big|h,1,\bar{h},0\rangle + i\frac{\bar{h}}{c}\big|h,1,\bar{h},0\rangle - \frac{i}{c}\big|h,0,\bar{h},0\rangle$$

$$= \xi|\Delta,\xi,1\rangle \qquad , \qquad (4.26)$$

or in short:

$$T^0|\Delta,\xi,0\rangle = \Delta|\Delta,\xi,0\rangle\ ,$$

$$T^0|\Delta,\xi,1\rangle = \Delta|\Delta,\xi,1\rangle + |\Delta,\xi,0\rangle\ ,$$

$$M^0|\Delta,\xi,0\rangle = \xi|\Delta,\xi,0\rangle\ ,$$

$$M^0|\Delta,\xi,1\rangle = \xi|\Delta,\xi,1\rangle\ . \qquad (4.27)$$

So, up to the order we would like to keep, Jordan form appears only for scaling and rapidity can't play a role. No matter if we have a holomorphic or antiholomorphic logarithmic form.

For fields based on state operator correspondence we have:

$$T^0\phi_{\Delta,\xi}(x,t)|0\rangle = \Delta\phi_{\Delta,\xi}(x,t)|0\rangle\ ,$$

$$T^0\psi_{\Delta,\xi}(z)|0\rangle = \Delta\psi_{\Delta,\xi}(z)|0\rangle + \phi_{\Delta,\xi}(z)|0\rangle\ . \qquad (4.28)$$

Now, for two-point functions we follow from contraction approach in the light of equations (2.8), (4.12):

$$\langle\phi_1(x_1,t_1)\phi_2(x_2,t_2)\rangle_{GCA} = 0$$

$$\langle\phi_1(x_1,t_1)\psi_2(x_2,t_2)\rangle_{GCA} = lim_{c\to\infty}\langle\phi_1(x_1,t_1)\psi_2(x_2,t_2)\rangle_{CFT}$$

$$= lim_{c\to\infty}\, b\delta_{h_1,h_2}\delta_{\bar{h}_1,\bar{h}_2}(t_{12} + i\ \frac{x_{12}}{c})^{-2h_1}(t_{12} - i\frac{x_{12}}{c})^{-2\bar{h}_1}$$

$$= b\delta_{\Delta_1,\Delta_2}\delta_{\xi_1,\xi_2}{t_{12}}^{-2\Delta_1}exp\ (\frac{2\xi_1 x_{12}}{t_{12}})\ . \qquad (4.29)$$

For last two-point function again we have:

$$\langle\psi_1(x_1,t_1)\psi_2(x_2,t_2)\rangle_{GCA} = lim_{c\to\infty}\langle\psi_1(x_1,t_1)\psi_2(x_2,t_2)\rangle_{CFT}$$

$$= \lim_{c\to\infty} z^{-2h}\bar{z}^{-2\bar{h}}(d - 2blog(z))$$

$$= \lim_{c\to\infty}(t + ix/c)^{-\Delta-ic\xi}(t - ix/c)^{-\Delta+ic\xi}(d - 2blog(t{+}\in x))$$

$$= \delta_{\Delta_1,\Delta_2}\delta_{\xi_1,\xi_2}t^{-2\Delta_1}\ exp\left(\frac{2\xi_1 x}{t}\right)(d - 2blog(t)). \qquad (4.30)$$

We, try to obtain a 3-function of the model as well:

$$\langle\phi_1(x_1,t_1)\psi_2(x_2,t_2)\psi_3(x_3,t_3)\rangle_{GCA=}\ lim_{c\to\infty}\langle\phi_1(x_1,t_1)\psi_2(x_2,t_2)\psi_3(x_3,t_3)\rangle_{CFT}$$

$$= \lim_{c\to\infty}(b - 2a\, log\left(t_{23} + \frac{ix_{23}}{c}\right))[( z_{12}{}^{h_3-h_1-h_2} z_{13}{}^{h_2-h_1-h_3} z_{23}{}^{h_1-h_3-h_2}) * C.C.]$$

$$= (b - 2a\, log\,(t_{23}))\lim_{c\to\infty}[( z_{12}{}^{h_3-h_1-h_2} z_{13}{}^{h_2-h_1-h_3} z_{23}{}^{h_1-h_3-h_2}) * C\,.C]. \quad (4.31)$$

So, for this 3-point function as well, logarithmic form appears only on the time direction.

All this procedure may be redone using the nilpotent variables [54], and directly looking at the non relativistic limit of the corresponding LCFT correlators; all the results of the above will be repeated.

## 5. Concluding remarks

In this paper we have shown that logarithmic terms may also appear in the NRCFT's. This may be relevant to non relativistic phenomena which exhibit conformal invariance. Though the non unitarity of LCFT is a point of concern, however in some statistical mechanical models where the probability interpretation of the norms is forgone this worry may be set aside. We also note that the logarithmic terms appear only in the time coordinate, and this persists for higher correlators as well. It must also be noted that a more sophisticated approach to LCFT via the staggered modules [55, 56] may yield interesting results. Work in this direction is in progress.

## 6-References

[1] J. M. Maldacena, "The large N limit of superconformal field theories and supergravity," Adv. Theor. Math. Phys. 2, 231 (1998) [Int. J. Theor. Phys. 38, 1113 (1999)] [hep-th/9711200].

[2] S. S. Gubser, I. R. Klebanov and A. M. Polyakov, "Gauge theory correlators from non-critical string theory," Phys. Lett. B 428, 105 (1998) [hep-th/9802109].

[3] E. Witten, "Anti-de Sitter space and holography," Adv. Theor. Math. Phys. 2, 253 (1998) [hep-th/9802150]

[4] D. T. Son, "Toward an AdS/cold atoms correspondence: a geometric realization of the Schrodinger symmetry," Phys. Rev. D 78, 046003 (2008) [hep-th/0804.3972 ].

[5] K. Balasubramanian and J. McGreevy, "Gravity duals for non-relativistic Conformal Field Theories," Phys. Rev. Lett. 101, 061601 (2008) [hep-th/0804.4053].

[6] S. A. Hartnoll, "Lectures on holographic methods for condensed matter physics," Class. Quant. Grav. 26, 224002 (2009) [hep-th/0903.3246].

[7] U. Niederer, "The maximal kinematical symmetry group of the free Schrodinger equation," Helv. Phys. Acta 45, 802 (1972).

[8] C. R. Hagen, “Scale and conformal transformations in Galilean-covariant field theory,” Phys. Rev. D 5, 377 (1972).

[9] R. Jackiw, “Introducing scaling symmetry,” Phys. Today, 25(1), 23 (1972).

[10] G. Burdet, M. Perrin, and P. Sorba, “About the non-relativistic structure of the conformal algebra,” Commun. Math. Phys. 34 , 85 (1973)

[11] W. Fushchych, V. Chopyk, P. Nattermann, W. Scherer, “Symmetries and reductions of nonlinear schrodinger equations of Doebner-Goldin type” Reports on Math. Phys., 35, 129 (1995)

[12] M. Alishahiha, R. Fareghbal, A. E. Mosaffa and S. Rouhani, “Asymptotic symmetry of geometries with Schrodinger isometry”, Phys. Lett. B 675, 133 (2009) [hep-th/0902.3916]

[13] G. Compère, S. de Buyl, S. Detournay and K. Yoshida, “Asymptotic symmetries of Schrödinger spacetimes”, JHEP, 0910, 032 (2009) [hep-th /0908.1402]

[14] A.A. Belavin, A.M. Polyakov and A.B. Zamolodchikov, “Infinite conformal symmetry in two-dimensional quantum field theory” Nucl. Phys. B 241, 333 (1984)

[15] P. Di Francesco, P. Mathieu, D. Senechal, Conformal Field Theory, Springer(1997).

[16] M. Henkel, “Schrodinger invariance in strongly anisotropic critical systems,” J Statist. Phys. 75, 1023 (1994) [hep-th/9310081].

[17] P. Havas and J. Plebański "Conformal extensions of the Galilei group and their relation to the Schrodinger group" J. Math. Phys. 19, 482 (1978).

[18] M. Henkel, “Phenomenology of local scale invariance: From conformal invariance to dynamical scaling,” Nucl. Phys. B 641, 405 (2002) [hep-th/0205256].

[19] M. Henkel “ Local scale invariance and strongly anisotropic equilibrium critical systems” Phys. Rev. Lett. 78, 1940 (1997), [cond-mat/9610174].

[20] A. Bagchi and R. Gopakumar, “Galilean conformal algebras and AdS/CFT,” JHEP, 0907, 037 (2009) [hep-th/0902.1385].

[21] A. Hosseiny, S. Rouhani, “Affine extension of Galilean conformal algebra in 2+1 dimensions”, J. Math. Phys. 51, 052307 (2010) [hep-th/0909.1203]

[22] S. Bhattacharyya, S. Minwalla and S. R. Wadia, “The incompressible non-relativistic Navier-Stokes equation from gravity,” JHEP 0908, 059 (2009) [hep-th/0810.1545].

[23] P.-M. Zhang, P.A. Horvathy, “Non Relativistic Conformal Symmetries in Fluid Mechanics”, Eur. Phys. J. C 65 607(2010) [physics.flu-dyn/0906.3594 ].

[24] R. Cherniha, M. Henkel, “The exotic conformal Galilei algebra and nonlinear partial differential equations,” J. Math. Anal. Appl. 369, 120 (2010) [hep-th/0910.4822].

[25] C. Duval and P. A. Horvathy, “The exotic Galilei group and the ``Peierls substitution",” Phys. Lett. B 479, 284 (2000) [hep-th/0002233].

[26] R. Jackiw and V. P. Nair, “Anyon spin and the exotic central extension of the planar Galilei group,” Phys. Lett. B 480, 237 (2000) [hep-th/0003130].

[27] J. Lukierski, P. C. Stichel and W. J. Zakrzewski, “Exotic Galilean Conformal Symmetry and Its Dynamical Realisations,” Phys. Lett. A 357, 1 (2006) 1[hep-th/0511259]

[28] J. Lukierski, P. C. Stichel, and W. J. Zakrzewski, "Acceleration-Extended Galilean Symmetries with Central Charges and their Dynamical Realizations," Phys. Lett. B 650, 203 (2007)

[29] J. Negro, M. A. del Olmo and A. Rodr´ıguez-Marco, “Nonrelativistic conformal groups. I,”J. Math. Phys. 38 (1997) 3786.

[30] C. Roger, J. Unterberger “The Schrödinger-Virasoro Lie group and algebra: from geometry to representation theory”, Ann. Henri Poincare 7, 1477 (2006) [math-ph/0601050]

[31] M. Henkel and J. Unterberger, “Schrodinger invariance and space-time symmetries,” Nucl. Phys. B 660, 407 (2003) [hep-th/0302187].

[32] M. Henkel, R. Schott, S. Stoimenov, and J. Unterberger, “The Poincare algebra in the context of ageing systems: Lie structure, representations, Appell systems and coherent states,” [math-ph/0601028]

[33] A. Bagchi and I. Mandal 2009, “On representations and correlation functions of Galilean conformal algebra”, Phys. Lett. B 675, 393 (2009) [hep-th/0903.4524]

[34] Arjun Bagchi, Rajesh Gopakumar, Ipsita Mandal, Akitsugu Miwa 2009, “GCA in 2d” [hep-th/0912.1090]

[35] M. Alishahiha, A. Davody and A. Vahedi, “On AdS/CFT of Galilean conformal field theories,” JHEP 0908, 022 (2009) [hep-th/0903.3953].

[36] D. Martelli and Y. Tachikawa, “Comments on Galilean conformal field theories and their geometric realization,” [hep-th/0903.5184].

[37] C. Duval and P. A. Horvathy, “Non-relativistic conformal symmetries and Newton-Cartan structures,” J. Phys. A 42, 465206 (2009) [math-ph/0904.0531].

[38] D. Giulini, “On Galilei invariance in quantum mechanics and the Bargmann superselection rule”, Ann. of Phys. 249, 222 (1996) [quant-ph/9508002]

[39] V. Gurarie, “Logarithmic operators in conformal field theory”, Nucl. Phys. B410, 535 (1993); [hep-th/9303160].

[40] M.R. Gaberdiel, H.G. Kausch, “A rational logarithmic conformal field theory”, Phys. Lett. B386, 131 (1996); [hep-th/9606050].

[41] M.R. Gaberdiel, H.G. Kausch, “A local logarithmic conformal field theory”, Nucl. Phys. B 538, 631 (1999); [hep-th/9807091].

[42] H. Saleur, “Polymers and percolation in two dimensions and twisted N = 2 supersymmetry”, Nucl. Phys. B382, 486 (1992); [hep-th/9111007].

[43] G.M.T. Watts, “A crossing probability for critical percolation in two dimensions”, J. Phys. A29, L363 (1996); [cond-mat/9603167].

[44] M.R. Rahimi Tabar, S. Rouhani, “A logarithmic conformal field theory solution for two dimensional magnetohydrodynamics in presence of the Alf ’ven effect”, Europhys. Lett. 37, 447 (1997); [hep-th/9606143].

[45] V. Gurarie, M. Flohr, C. Nayak, “The Haldane-Rezayi quantum hall state and conformal field theory”, Nucl. Phys. B498, 513 (1997); [cond-math/9701212].

[46] S. Mathieu, P. Ruelle, “Scaling fields in the two-dimensional abelian sandpile model”, Phys. Rev. E 64, 066130 (2001) [hep-th/0107150].

[47] J.-S. Caux, I.I. Kogan, A.M. Tsvelik, “Logarithmic operators and hidden continuous symmetry in critical disordered models”, Nucl. Phys. B466, 444 (1996); [hep-th/ 9511134].

[48] M. Flohr, “Bits and pieces in logarithmic conformal field theory”, Int. J. Mod. Phys. A18 (2003) 4497–4592 [hep-th/0111228].

[49] M. R. Gaberdiel, “An algebraic approach to logarithmic conformal field theory”, Int. J. Mod. Phys. A18 (2003) 4593–4638 [hep-th/0111260].

[50] Daniel Grumiller, Olaf Hohm, “AdS3/LCFT2 – correlators in new massive gravity”, Phys. Lett. B 686, 264 (2010) [hep-th/0911.4274].

[51] D. Grumiller, I. Sachs, ”AdS3/LCFT2 – correlators in cosmological topologically massive gravity”, JHEP 1003, 012 (2010) [hep-th/0910.5241].

[52] K. Skenderis, M. Taylor and B. C. van Rees,”Topologically massive gravity and the AdS/CFT correspondence”, JHEP 0909, 045 (2009) [hep-th/0906.4926].

[53] Y. Nakayama, ”An Index for Non-relativistic Superconformal Field Theories “ JHEP 0810:083,(2008), [hep-th/0807.3344]

[54] S Moghimi-Araghi, S Rouhani, M Saadat, “Logarithmic conformal field theory through nilpotent conformal dimension”, Nucl. Phys. B, 599, 531 (2000) [hep-th/0008165].

[55] F Rohsiepe “On reducible but indecomposable representations of the Virasoro algebra”, [hep-th/9611160].

[56] P Mathieu, D Ridout “From percolation to logarithmic conformal field theory”, Physics Letters B, 657, 120 (2007), [hep-th/0708.0802]